# Spectral-change enhancement with prior SNR for the hearing impaired

Xiang LI[1]; Xin TIAN[2]; Henry LUO[3]; Jinyu QIAN[3]; Xihong WU[1]; Dingsheng LUO[1]; Jing CHEN[1]

[1] Peking University, China

[2] Sonova, China

[3] Sonova, Canada

**ABSTRACT**

A previous signal processing algorithm that aimed to enhance spectral changes (SCE) over time showed benefit for hearing-impaired (HI) listeners to recognize speech in background noise. In this work, the previous SCE was manipulated to perform on target-dominant segments, rather than treating all frames equally. Instantaneous signal-to-noise ratios (SNRs) were calculated to determine whether the segments should be processed. Initially, the ideal SNR calculated by the knowledge of premixed signals was introduced to the previous SCE algorithm (SCE-iSNR). Speech intelligibility (SI) and clarity preference were measured for 12 HI listeners in steady speech-spectrum noise (SSN) and six-talk speech (STS) maskers, respectively. The results showed the SCE-iSNR algorithm improved SI significantly for both maskers at high signal-to-masker ratios (SMRs) and for STS masker at low SMRs, while processing effect on speech quality was small. Secondly, the estimated SNR obtained from real mixtures was used, resulting in another SCE-eSNR. SI and subjective rating on naturalness and speech quality were tested for 7 HI subjects. The SCE-eSNR algorithm showed improved SI for SSN masker at high SMRs and for STS masker at low SMRs, as well as better naturalness and speech quality for STS masker. The limitations of applying the algorithms are discussed.

Keywords: Spectral-change enhancement, Hearing impairment, Prior SNR, Speech intelligibility, Speech quality

## 1. INTRODUCTION

Hearing-impaired listeners can usually achieve high speech intelligibility in a quiet environment, but their speech understanding deteriorates in noisy backgrounds. Most people with sensorineural hearing loss is associated with broader-than-normal auditory filters (1). The excitation pattern becomes "smeared" compared with the normal hearing when the speech signal passes through the broader filters. To be specific, the difference between peaks and dips in amplitude spectrum is reduced, resulting in the poor representation of some important perceptual cues, e.g. formant frequencies (2). Background noise then fills the dips in spectrum, which further reduces spectral contrast in the signal, exacerbating the problem of perceiving spectral cues for people with hearing loss.

To compensate for this effect, most attempts mainly focused on enhancing spectral contrast (3). However, these studies only enhanced spectral contrast in individual frames while ignored the spectral changes across frames. Most information in speech is actually carried in the spectral changes over time, rather than in static spectral shape per se (4). Several studies have suggested that suitable enhancement on spectral changes could improve speech intelligibility for hearing impaired people, especially in noisy environments. In previous studies (5), a signal processing algorithm has been realized by evaluating spectral change across adjacent temporal frames and enhancing the spectral changes with personalized parameter settings selected using a genetic algorithm. The algorithm was evaluated by measuring speech intelligibility and clarity preference for hearing-impaired (HI) listeners. The benefit of this algorithm was significant but small.

To extend the benefit of the above mentioned spectral-change enhancement (SCE) algorithm, the current study was to enhance the spectral change mostly for the target speech, but not for the interfering sounds. This can be achieved by first estimating the instantaneous signal-to-noise ratios (SNRs) for each frame. Then, the SNRs were utilized to determine whether the spectral change was dominated by the target or by the masker for a mixture. The prior SNR has been widely used in speech enhancement algorithms and algorithms for SNR evaluation have been widely implemented in hearing aids (6). In the previous algorithm, SCE was operated equally across temporal frames. Differently in this work, the SNR information of each frame would be used to determine when the SCE should be applied so as to avoid the destructive effect of enhancing spectral changes of the maskers.

---

[1] janechenjing@pku.edu.cn

In this paper, the previous SCE algorithm was modified by introducing instantaneous SNR information, so that the algorithm could selectively process the frames with SNRs above a given SNR threshold. To separate the quality of the SNR estimation from the effect of the modified algorithm itself, the ideal SNR (iSNR) was first applied. The iSNR was calculated from the knowledge of premixed signals. This new iSNR based SCE algorithm was referred to as the SCE-iSNR, which was evaluated via experiments of speech intelligibility and clarity preference using speech with steady speech-spectrum noise (SSN) and six-talk speech (STS) maskers for 12 hearing-impaired listeners. Secondly, since the iSNR cannot be available in real-world scenarios, the estimated SNR (eSNR) from real mixtures was applied to replace the iSNR, resulting in another new algorithm, the SCE-eSNR. Additional experiments evaluated the speech intelligibility and subjective rating of naturalness and speech quality for 7 HI subjects.

The remainder of the paper was organized as follows. We first introduced the previous SCE algorithm briefly and then the application of the instantaneous SNR, including the iSNR and the eSNR. The proposed SCE-iSNR and SCE-eSNR were evaluated separately in terms of speech intelligibility and subjective speech quality in the follow-up sections.

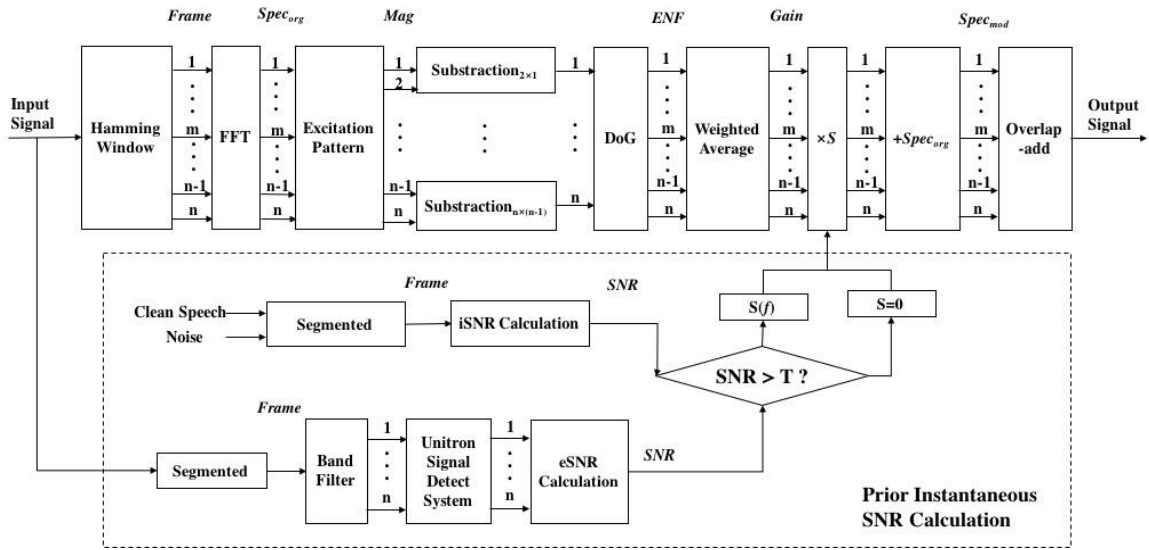

Figure 1 – Flow diagram of the processing for spectral-change enhancement with instantaneous SNR.

## 2. SIGNAL PROCESSING

The modified signal processing algorithms, SCE-iSNR and SCE-eSNR were divided into two parts: SCE processing and instantaneous SNR calculation. For the SCE processing, the input signal was segmented, windowed, spectrally smoothed, spectral-change enhanced, and then resynthesized with the overlap-add technique. For instantaneous SNR calculation, two methods were considered, including iSNR calculation and eSNR calculation. These two parts were combined by the equation (1)

$$S_n = \begin{cases} S, & SNR_n \geq T \\ 0, & SNR_n < T \end{cases} \qquad (1)$$

where $S_n$ is a parameter that controls the degree of the enhancement in frame $n$, $SNR_n$ represents the SNR of frame $n$, $T$ represents the SNR threshold which is set as 0 dB in our experiments. According to equation (1), when the frame SNR is below the threshold, the SCE degree is 0, suggesting no SCE was operated in this frame. The processing flow is shown in Figure 1.

## 2.1 SCE Processing

Since the processing has been described in details in the earlier paper (5), a simply outline of steps are provided as the following.

The input signal was first segmented and Hamming windowed, followed by a 256-point FFT, giving 128 original magnitude values ($Spec_{org}$) and phase values. A convolution procedure (7) was used to transform the original magnitude spectrum to an auditory excitation pattern, denoted as $Mag$. The initial spectral change was obtained by calculating the ratio of the $Mag$ value across every two adjacent frames. An enhancement function ($ENF$) was then obtained by convoluting the initial spectral change with a Difference of Gaussian ($DoG$) function.

To obtain the modified magnitude spectrum ($Spec_{mod}$) for a given frame, a Gain function was added to the $Spec_{org}$. This Gain function for a certain frame $n$, denoted as $Gain_n$, was defined by a weighted (weight $\xi<1$) average of the $ENF$ over $m$ preceding frames. Then, the $Gain_n$ was multiplied by a factor $S$, which was an adjustable parameter used to control the degree of spectral-change enhancement. Finally, the $Spec_{mod}$ of each frame was obtained by adding the scaled $Gain_n$ to the $Spec_{org}$. The corresponding output signal was resynthesized by IFFT with the $Spec_{mod}$ and its original phase.

In summary, there are four manipulated parameters to implement the whole processing: $b$, the width of the DoG function specified in $ERB_N$ number; $\xi$ and $m$, controlling the effect of preceding frames; and $S$, controlling the amount of enhancement.

## 2.2 Instantaneous SNR Calculation

Two methods for instantaneous SNR calculation were considered. The first one was based on *iSNR calculation* which was obtained from premixed signals. For the application in real-world scenarios, the second one was based on *eSNR calculation* which was obtained from real mixed signals. The resulting SNR information was used to manipulate the parameter $S$ of SCE processing by equation (1).

1) *iSNR calculation:* The target signal and the masker signal were segmented frame by frame separately in the same way as the SCE processing. Then, the iSNR for each frame was calculated by the ratio of the target signal energy and the masker signal energy of each frame.
2) *eSNR calculation:* Noisy speech from real mixtures was measured frame by frame using the real-time SNR detector of Unitron hearing aid (8). To be specific, an input signal was first segmented and divided into a set of frequency domain input signals. Each frequency domain signal was then analyzed based on three characteristics: intensity change, modulation frequency and time duration, each of which produced a sub-index. Three sub-indexes were further combined together to produce a signal index using a three-dimensional continuum. According to the signal index, the frequency domain input signal was classified as different types of noise or as a desired signal, which was used to produce SNR values of each frequency band. The final eSNR for each frame was calculated by averaging the SNR values across frequency bands.

## 3. GENERAL EXPERIMENT METHOD

In the previous study (5), a genetic algorithm (GA) was used to select the best parameter setting when implementing SCE for each subject. Subjects should compare the subjective intelligibility of successive sentences processed by SCE with different parameter values in background noise. It is necessary to select an appropriate sentence SMR to avoid too high or too low for the GA test, thus a speech reception threshold (SRT) was measured for unprocessed stimuli before the GA test. The SRT value was also used to determine the high SMR and low SMR for the following tests of speech intelligibility (SI) and speech quality, including clarity preference (CP) and subjective rating (SR).

Two experiments were conducted to evaluate the effect of SCE-iSNR and SCE-eSNR, respectively. Experiment I was designed to evaluate the SCE-iSNR relative to the previous SCE and unprocessed condition, while Experiment II was designed to evaluate the SCE-eSNR relative to the speech enhancement algorithm installed in Unitron hearing aids (SE-Unitron) and unprocessed condition.

In general, the following measurements need to be determined in the experiments: 1) SRT measurement; 2) GA test; 3) SI measurement; 4) CP measurement; 5) SR measurement. Therein, 1) - 3) were measured for both Experiment I and Experiment II while 4) was only measured for Experiment I and 5) was only measured for Experiment II.

### 3.1 Speech Material

Sentences from the Mandarin Hearing in Noise Test (MHINT) database (9) were used as the target

speech for the SRT and SI measurements. Sentences from Chinese nonsense sentences corpus (10) were used as the target speech for the GA and CP measurements. Two speech corpuses were used because the sentence amount of each existing Mandarin speech corpus was too small to cover all tests.

The SSN masker had the same spectral shape as the averaged target speech for the corresponding tests. STS masker was from a 10-s sample of six-talker Mandarin Chinese babble (11). For each trial, the masker started 500 ms before the target speech, and finished synchronously with the target.

### 3.2 Speech Reception Threshold (SRT)

The SRT was measured for SSN and STS masker, respectively, each using 2 lists (20 sentences) from MHINT database. The initial SMR was selected to be relatively high to ensure most key words to be heard at the beginning. If more than half key words were repeated correctly, the SMR was decreased by 4 dB; otherwise the SMR was increased by 4 dB. After four sentences had been presented, the step size in SMR was set to 2 dB. The SRT for each condition was determined by averaging the SMRs at the last four turn points.

### 3.3 Genetic Algorithm (GA)

The GA tool developed by (12) was used to select the "best" parameter values for SCE, SCE-iSNR in Experiment I and for SCE-eSNR in Experiment II, respectively for each subject. The test procedure was the same as that in the previous study (5), so the detailed description was omitted here.

The GA was run separately for each of the testing conditions (2 maskers × 2 processing types for Experiment I; 2 maskers × 1 processing types for Experiment II). The selected parameter values were then used to process the sentences in corresponding conditions for SI, CP and SR measurements.

### 3.4 Measurement of Speech Intelligibility (SI)

Two SMRs were selected for each masker based on the SRT results for each subject: SMR_H (1 dB above the SRT), and SMR_L (2 dB below the SRT). There were twelve conditions: 2 masker types × 2 SMRs × 3 processing types. Subjects were instructed to vocally repeat the ten key words of each sentence. The experimenter recorded the key words that had been identified correctly.

### 3.5 Measurement of Clarity Preference (CP)

In each trial, three speech stimuli representing three processing types were presented with 500-ms inter-stimulus-interval time. The same sentence was used for each trial, and three processing types were assigned randomly to the three intervals. Subjects should make a choice among the three intervals following the instruction: which sentence of the intervals is the best of clarity for you? The order of four conditions (2 masker types × 2 SMRs) was balanced across subjects using a 4-order Latin-square and four lists (18 sentences per list) were randomly assigned to the four conditions.

### 3.6 Measurement of Subjective Rating (SR)

Double-blinded Multiple Stimuli with Hidden Reference and Anchor (MUSHRA) test was used to access the subjective rating of naturalness and overall speech quality separately among three speech samples obtained from one sentence processed by three processing types for each trial. The order of four conditions (2 masker types × 2 SMRs) was random across subjects and each condition contained 4 sentences randomly selected from four lists (18 sentences per list). Subjects listened to each sentence and then were asked to indicate their rating score based on a given rating scale.

## 4. EXPERIMENT I

### 4.1 Experimental Procedure

The experimental procedure for Experiment I was designed as follows: SRT was first measured, and then the GA test, followed by SI measurement and finally CP measurement.

The GA test was run separately for SCE and SCE-iSNR with two maskers for each subject. Four parameters need to be selected: $b$, $\xi$, $m$, $S$. For each parameter, minimums were 0.5, 0.8, 5, 1 and maximums were 3, 0.9, 6, 5 with step size of 0.5, 0.1, 1, 0.5, respectively. These were chosen to avoid obvious audio distortion from SCE processing and to ensure the effective convergence of GA tool.

In the SI measurement, the order of the twelve conditions was balanced across subjects using a Latin-square. Twelve lists of MHINT sentences were assigned randomly to these twelve conditions for each subject. SRT and CP were measured as described in section 3.2 and 3.5, respectively.

### 4.2 Subjects and Compensation for hearing loss

Twelve subjects with moderate to severe hearing loss were tested (without hearing aids). Stimuli were presented to the better ear which was determined by the average audiometric threshold from 0.25 to 6 kHz. To compensate for the hearing loss, a linear amplification according to the Cambridge formula (13) was applied to stimuli before they were presented to subjects.

### 4.3 Stimuli

All signals were presented with an external soundcard via Sennheiser HD 650 headphones. The sound level was calibrated at 65 dB SPL for the target speech prior to the amplification prescribed by the Cambridge formula. Subjects were seated in a double-walled sound attenuating chamber.

### 4.4 Results

#### 4.4.1 Speech Reception Threshold (SRT)

Individual SRTs for unprocessed stimuli for each masker were plotted as a function of the subject's mean audiometric threshold (MAT) in Figure 2. The SRTs varied markedly across subjects. For each subject, the SRT was lower for the SSN than for the STS masker, except for the subject 11. The SRTs were significantly correlated with the MAT for both SSN (r=0.828, p=0.001) and STS (r=0.703, p=0.011) makers. In general, the higher MAT, the poorer was the ability to understand speech in background noise.

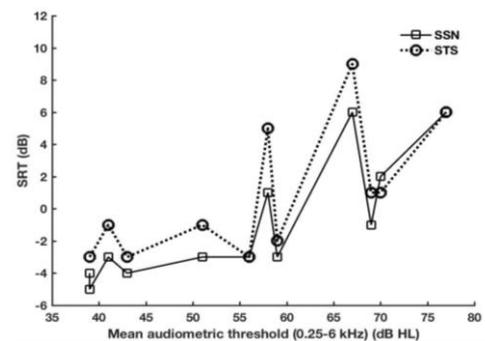

Figure 2 – SRTs for unprocessed stimuli for each subject, ordered according to the mean audiometric thresholds.

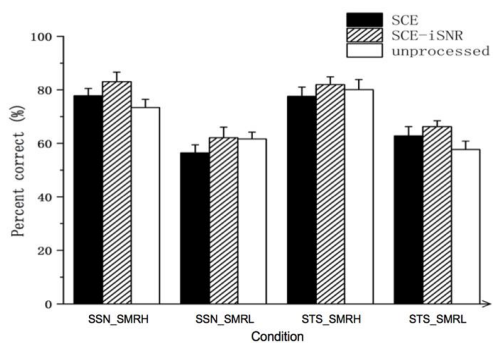

Figure 3 – Mean percent correct identification of key words across 12 subjects for each test condition. Error bars indicate ±1 standard error.

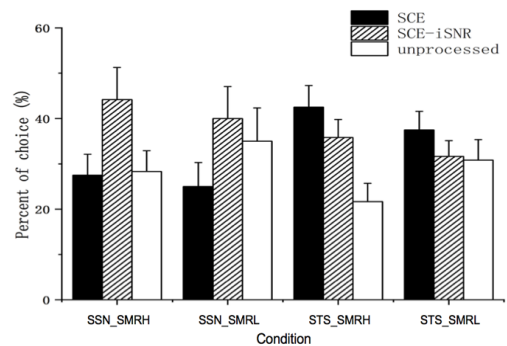

Figure 4 – Mean percentage of selections of each response category across 12 subjects for each test condition. Error bars indicate ±1 standard error.

#### 4.4.2 Speech Intelligibility (SI)

Figure 3 showed the mean speech intelligibility scores across 12 subjects. The results are presented in four groups according to the masker type and SMR, with average correct key words percent for each of the three processing types within each group.

In general, the performance became better with the increase of SMR for both maskers. The SCE-iSNR processing achieved the highest average scores across all conditions. A three-way (SMR, masker type, and processing type) analysis of variance (ANOVA) indicated that main effect was significant for SMR [$F(1,11)=122.410$, $p<0.001$] and processing type [$F(2,22)=8.310$, $p=0.002$], but not significant for masker type. The interaction across the three factors was significant [$F(2,22)=8.311$, $p=0.002$]. The post hoc Bonferroni pairwise comparison showed significantly higher SI scores for SCE-iSNR processing than for both SCE processing [$p=0.034$] and unprocessed condition [$p=0.014$]. Furthermore, the SI score was significantly higher for SCE-iSNR than for unprocessed condition in the groups of SSN masker at high SMR [$p=0.026$] and STS masker at low SMR [$p=0.009$]. Overall, HI listeners had better average speech intelligibility for the SCE-iSNR processing compared with the unprocessed condition.

### 4.4.3 Clarity Preference (CP)

Figure 4 showed the mean subjective clarity preference across the 12 subjects. For each group in the figure, the summation of the percentage of the three options was 100%, hence the later statistic analysis was based on the percentage for the two processing conditions excluding the condition of "unprocessed". It was clearly that SCE-iSNR processing was chosen more often than the other two processing types for SSN masker, while the SCE processing was selected the most for STS masker. A three-way analysis of variance (ANOVA) was conducted with two masker types, SMR and two processing types (SCE and SCE-iSNR). The main effect of SMR [$F(1,11)=4.945$, $p=0.048$] and the interaction between masker type and processing type [$F(1,11)=4.962$, $p=0.048$] were significant. Then, a two-way ANOVAs was conducted for SMR_H and SMR_L, respectively, using masker type and processing type as the two factors. No any main effects and interactions was significant for the analysis. We can roughly conclude there was no significant perceptual difference between each pair of processing types for hearing-impaired subjects on average.

## 5. EXPERIMENT II

### 5.1 Experimental Procedure

The experimental procedure for Experiment II was designed as follows: SRT was first measured, and then GA test was run, followed by SI measurement and finally SR measurement.

The GA tool was run only for SCE-eSNR processing with two maskers for each subject here. To reduce the workload of labeling eSNRs during the eSNR calculation, the values of $\xi$ and $m$ were fixed here as 0.9 and 5 according to the observation of previous experiment results (5). The minimum, maximum and step values for parameters $b$ and $S$ were the same as those in Experiment I.

In the SI measurement, the order of the twelve conditions was random across 7 subjects. Twelve lists of MHINT sentences were assigned randomly to these twelve conditions for each subject. SRT and SR were measured as descripted in section 3.2 and 3.6, respectively.

### 5.2 Subjects and Compensation for hearing loss

Experiment II was conducted by Sonova R&D China for seven subjects with symmetrical mild to severe hearing loss, defined as an interaural difference of no more than 10 dB across the octave frequencies from 250 to 8000 Hz in audiometric thresholds. Stimuli were presented binaurally here. To compensate for the hearing loss, all subjects were initial binaurally fitted to hearing aids (Unitron North Stride Pro P BTE) with customized earmold according to the default first fit while a NAL-NL2 Tonal fitting formula was used.

### 5.3 Stimuli

The presented stimuli which was the recording of streamed hearing aid output, together with eSNR estimation were created synchronously using two Unitron North Stride Pro P recording hearing aids. The signal was recorded with a RME Fireface 802 sound card and Sound Forge Pro 11.0. The stimuli were presented from a loudspeaker located 1.4 meters in front of subjects in a free field. The calibration of speech material was created by merging all the sentences in advance. The volume was adjusted until the values met 65 dB SPL for the most of middle frequency bands.

### 5.4 RESULTS

### 5.4.1 Speech Reception Threshold (SRT)

In Figure 5, similar to the result in Experiment I, the SRT was lower for the SSN than for the STS masker, except for the subject 6. There was no significant correlation between SRTs and the MAT for either SSN or STS masker, which might be caused by subject 1 whose MAT was the lowest (53 dB HL) but achieving a relatively high SRT value for STS masker, and by subject 7 whose MAT was the highest (even more than 70 dB HL) but his SRT values were the lowest for both maskers.

### 5.4.2 Speech Intelligibility (SI)

The performance became better with the increase of SMR for both maskers in Figure 6. A three-way (SMR, masker type, and processing type) analysis of variance (ANOVA) indicated that the main effect of SMR was significant [$F(1,6)=78.913$, $p<0.001$] while no other main effect or interactions was significant. Due to the limit of subject amount, no further statistic analysis was conducted. However, some trends were observed as specified below. The SCE-eSNR processing showed a similar performance with the SE-Unitron for STS masker at high SMR. In the condition of

SSN masker at low SMR, the SCE-eSNR resulted in the highest average scores among three processing types. The SE-Unitron showed more benefit in the other two conditions, SSN with high SMR and STS with low SMR. Another observation was that both SE-Unitron and SCE-eSNR reduced SI for STS maskers at high SMR.

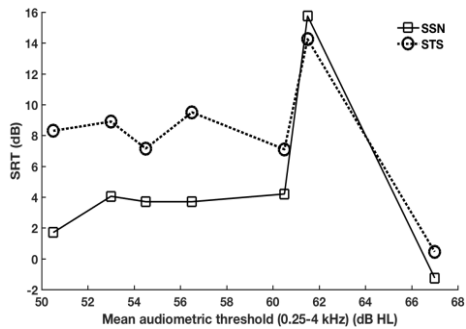 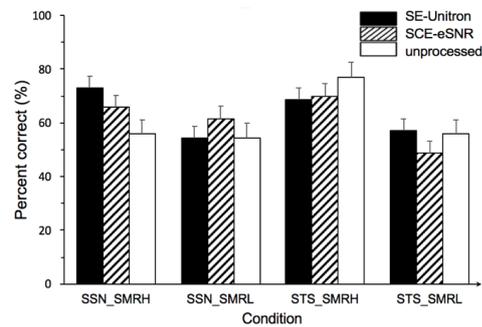

Figure 5 – SRTs for unprocessed stimuli for each subject, ordered according to the mean audiometric thresholds.

Figure 6 – Mean percent correct identification of key words across 7 subjects for each test condition. Error bars indicate ±1 standard error.

### 5.4.3 Subjective Rating (SR)

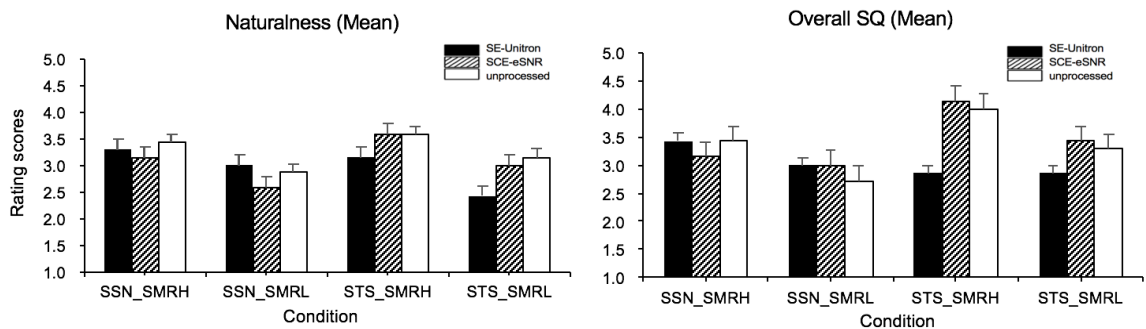

Figure 7 – Mean scores for the ratings of naturalness and speech quality across 7 subjects for each test condition. Error bars indicate ±1 standard error.

Figure 7 showed the mean subjective ratings of naturalness (NAT) and speech quality (SQ) across 7 subjects. Generally, SE-Unitron processing showed more benefit for SSN masker while SCE-eSNR processing reflected its advantage on STS masker.

A three-way (SMR, masker type, and processing type) analysis of variance (ANOVA) was conducted for rating results of NAT and SQ, respectively. For NAT, neither main effects nor interactions were significant. For SQ, main effect of SMR [$F(1,6)=11.267$, $p=0.015$] and interaction across three factors [$F(2,12)=6.783$, $p=0.011$] were significant. The post hoc Bonferroni pairwise comparison showed a significantly but slightly higher score for SCE-eSNR processing than for SE-Unitron in the condition of STS masker at low SMR [$p=0.050$]. There was an overall trend that SCE-eSNR processing produced a better speech quality for STS masker especially in low SMR.

## 6. DISCUSSION AND CONCLUSION

The previous SCE algorithm was modified by introducing the instantaneous SNR information. The effect of this modification was evaluated for the iSNR and eSNR separately by the speech intelligibility (SI) and subjective speech quality experiments for hearing-impaired listeners.

The SCE-iSNR processing resulted in improved SIs for both SSN and STS maskers at high SMR and for STS masker at low SMR compared to the unprocessed condition. These results were statistically significant. For the SCE-eSNR processing, although not significant, a trend of improved SI for SSN masker at low SMRs was observed compared to the unprocessed condition. The less clear benefit of the SCE-eSNR might be related to the less accuracy of the instantaneous SNR (eSNR here) or out of sync in time frame. The eSNR was obtained by a real-time SNR estimation algorithm, where the bias was naturally existed compared with the ground truth (e.g., iSNR). Non-stationary background noise (e.g., STS) with high energy (e.g., low SMRs) might lead to less accuracy of the eSNR, then resulting in a degraded effect for SCE-eSNR processing. Therefore, the SCE-eSNR for STS masker at both high and low SMRs even reduced the speech intelligibility (see Figure 6).

The effect of SCE-iSNR processing on speech quality was relatively small since no significant difference was observed between the SCE-iSNR and unprocessed condition. It was encouraging that SCE-eSNR showed an advantage on subjective rating for STS masker while SE-Unitron showed more benefit for SSN masker.

## ACKNOWLEDGEMENTS

This work was supported by a research funding by the SONOVA hearing technology company, China, and the National Natural Science Foundation of China (Grant Nos. 61473008, 61771023).